\documentclass[11pt]{article}
\usepackage{graphicx}

\newcommand{\BABARPubYear}    {05}

\newcommand{\BABARProcNumber} {35}
\newcommand{\SLACPubNumber} {11463}

\def\BtoXsll     {\ensuremath{\B \to X_s\: \ell^+ \ell^-}}

\def\modekavgll {\ensuremath{\B\to K\ellell}\xspace}
\def\modekstll {\ensuremath{\B\rightarrow K^{*}\ellell}\xspace}
\def\modekll {\ensuremath{\Bp \rightarrow K^+\ellell}\xspace}
\def\modekstll {\ensuremath{\B\rightarrow K^{*}\ellell}\xspace}
\newcommand{\mkpipi}{\ensuremath{m_{K\pi\pi}}\xspace}
\newcommand{\btaunu}     {\ensuremath{\Bp \to \taup \nut}\xspace}
\newcommand{\btodszlnu} {\ensuremath {\B^- \to D^{*0} \ell^- \nulb}}
\newcommand{\BRbtaunu}   {\ensuremath{\BR(\Bp \to \taup \nut)}\xspace}
\newcommand{\ccc}{\ensuremath{\Kp\pim\pip\gamma}\xspace}
\newcommand{\ccn}{\ensuremath{\Kp\pim\piz\gamma}\xspace}
\def\bkog    {\ensuremath {\Bz \to \Kstarz \gamma}}
\def\bkpg    {\ensuremath {\Bp \to \Kstarp \gamma}}

\RequirePackage{xspace}

\def\lbabar{\mbox{{\large\sl B}\hspace{-0.4em} {\normalsize\sl A}\hspace{-0.03em}{\large\sl B}\hspace{-0.4em} {\normalsize\sl A\hspace{-0.02em}R}}}
\usepackage{relsize}
\def\babar{\mbox{\slshape B\kern-0.1em{\smaller A}\kern-0.1em
    B\kern-0.1em{\smaller A\kern-0.2em R}}}

\def\taup       {\ensuremath{\tau^+}\xspace}

\def\ellell     {\ensuremath{\ell^+ \ell^-}\xspace}

\def\nub        {\ensuremath{\overline{\nu}}\xspace}

\def\nub        {\ensuremath{\overline{\nu}}\xspace}

\def\nut        {\ensuremath{\nu_\tau}\xspace}

\def\nulb       {\ensuremath{\nub_\ell}\xspace}

\def\g     {\ensuremath{\gamma}\xspace}

\def\W      {\ensuremath{W}\xspace}
\def\Wp     {\ensuremath{W^+}\xspace}

\def\u     {\ensuremath{u}\xspace}

\def\d     {\ensuremath{d}\xspace}

\def\sbar  {\ensuremath{\overline s}\xspace}

\def\b     {\ensuremath{b}\xspace}
\def\bbar  {\ensuremath{\overline b}\xspace}

\def\piz   {\ensuremath{\pi^0}\xspace}

\def\pip   {\ensuremath{\pi^+}\xspace}
\def\pim   {\ensuremath{\pi^-}\xspace}

\def\Kbar  {\kern 0.2em\overline{\kern -0.2em K}{}\xspace}

\def\Kz    {\ensuremath{K^0}\xspace}
\def\Kzb   {\ensuremath{\Kbar^0}\xspace}
\def\KzKzb {\ensuremath{\Kz \kern -0.16em \Kzb}\xspace}
\def\Kp    {\ensuremath{K^+}\xspace}
\def\Km    {\ensuremath{K^-}\xspace}

\def\KpKm  {\ensuremath{\Kp \kern -0.16em \Km}\xspace}

\def\Kstarz  {\ensuremath{K^{*0}}\xspace}

\def\Kstarp  {\ensuremath{K^{*+}}\xspace}

\def\Dbar    {\kern 0.2em\overline{\kern -0.2em D}{}\xspace}

\def\Dz      {\ensuremath{D^0}\xspace}
\def\Dzb     {\ensuremath{\Dbar^0}\xspace}
\def\DzDzb   {\ensuremath{\Dz {\kern -0.16em \Dzb}}\xspace}
\def\Dp      {\ensuremath{D^+}\xspace}
\def\Dm      {\ensuremath{D^-}\xspace}

\def\DpDm    {\ensuremath{\Dp {\kern -0.16em \Dm}}\xspace}

\def\B       {\ensuremath{B}\xspace}
\def\Bbar    {\kern 0.18em\overline{\kern -0.18em B}{}\xspace}

\def\BB      {\ensuremath{B\Bbar}\xspace} 
\def\Bz      {\ensuremath{B^0}\xspace}
\def\Bzb     {\ensuremath{\Bbar^0}\xspace}
\def\BzBzb   {\ensuremath{\Bz {\kern -0.16em \Bzb}}\xspace}
\def\Bu      {\ensuremath{B^+}\xspace}
\def\Bub     {\ensuremath{B^-}\xspace}
\def\Bp      {\ensuremath{\Bu}\xspace}
\def\Bm      {\ensuremath{\Bub}\xspace}

\def\BpBm    {\ensuremath{\Bu {\kern -0.16em \Bub}}\xspace}

\mathchardef\Upsilon="7107
\def\Y#1S{\ensuremath{\Upsilon{(#1S)}}\xspace}

\def\FourS {\Y4S}

\mathchardef\Deltares="7101
\mathchardef\Xi="7104
\mathchardef\Lambda="7103
\mathchardef\Sigma="7106
\mathchardef\Omega="710A

\def\Deltabar{\kern 0.25em\overline{\kern -0.25em \Deltares}{}\xspace}
\def\Lbar{\kern 0.2em\overline{\kern -0.2em\Lambda\kern 0.05em}\kern-0.05em{}\xspace}
\def\Sigbar{\kern 0.2em\overline{\kern -0.2em \Sigma}{}\xspace}
\def\Xibar{\kern 0.2em\overline{\kern -0.2em \Xi}{}\xspace}
\def\Obar{\kern 0.2em\overline{\kern -0.2em \Omega}{}\xspace}
\def\Nbar{\kern 0.2em\overline{\kern -0.2em N}{}\xspace}
\def\Xb{\kern 0.2em\overline{\kern -0.2em X}{}\xspace}

\def\BR         {{\ensuremath{\cal B}\xspace}}

\newcommand{\tev}{\ensuremath{\mathrm{\,Te\kern -0.1em V}}\xspace}
\newcommand{\gev}{\ensuremath{\mathrm{\,Ge\kern -0.1em V}}\xspace}
\newcommand{\mev}{\ensuremath{\mathrm{\,Me\kern -0.1em V}}\xspace}
\newcommand{\kev}{\ensuremath{\mathrm{\,ke\kern -0.1em V}}\xspace}
\newcommand{\ev}{\ensuremath{\mathrm{\,e\kern -0.1em V}}\xspace}
\newcommand{\gevc}{\ensuremath{{\mathrm{\,Ge\kern -0.1em V\!/}c}}\xspace}
\newcommand{\mevc}{\ensuremath{{\mathrm{\,Me\kern -0.1em V\!/}c}}\xspace}
\newcommand{\gevcc}{\ensuremath{{\mathrm{\,Ge\kern -0.1em V\!/}c^2}}\xspace}
\newcommand{\mevcc}{\ensuremath{{\mathrm{\,Me\kern -0.1em V\!/}c^2}}\xspace}

\def\mus  {\ensuremath{\rm \,\mus}\xspace}

\def\mus        {\ensuremath{\,\mu{\rm s}}\xspace}    

\def\to                 {\ensuremath{\rightarrow}\xspace}

\def\pep2{PEP-II}

\def\gsim{{~\raise.15em\hbox{$>$}\kern-.85em
          \lower.35em\hbox{$\sim$}~}\xspace}
\def\lsim{{~\raise.15em\hbox{$<$}\kern-.85em
          \lower.35em\hbox{$\sim$}~}\xspace}

\def\CP                {\ensuremath{C\!P}\xspace}

\def\Vub  {\ensuremath{|V_{ub}|}\xspace}

\xspace

\newcommand{\epjBase}        {Eur.\ Phys.\ Jour.\xspace}

\newcommand{\jprBase}        {Phys.\ Rev.\xspace}
\newcommand{\jplBase}        {Phys.\ Lett.\xspace}
\newcommand{\nimBaseA}       {Nucl.\ Instr.\ Meth.\xspace}

\newcommand{\nimBaseC}       {Nucl.\ Instr.\ and Methods\xspace}

\newcommand{\npBase}         {Nucl.\ Phys.\xspace}
\newcommand{\zpBase}         {Z.\ Phys.\xspace}

\newcommand{\epjc}      [1]  {\epjBase\ C~{\bf #1}}

\newcommand{\mpl}       [1]  {{Mod.\ Phys.\ Lett.\ {\bf #1}}}

\newcommand{\nim}       [1]  {\nimBaseC~{\bf #1}}
\newcommand{\nima}      [1]  {\nimBaseA~A~{\bf #1}}

\newcommand{\npb}       [1]  {\npBase\ B~{\bf #1}}

\newcommand{\npbps}     [1]  {{Nucl.\ Phys.\ B~Proc.\ Suppl.\ {\bf #1}}}

\newcommand{\plb}       [1]  {\jplBase\ B~{\bf #1}}

\newcommand{\pr}        [1]  {\jprBase\ {\bf #1}}
\newcommand{\jprd}      [1]  {\jprBase\ D~{\bf #1}}

\newcommand{\progtp}    [1]  {{Prog.\ Th.\ Phys.\ {\bf #1}}}

\newcommand{\jrmp}      [1]  {{Rev.\ Mod.\ Phys.\ {\bf #1}}}  

\newcommand{\zpc}       [1]  {\zpBase\ C~{\bf #1}}

\def\jetset74   {\mbox{\tt Jetset \hspace{-0.5em}7.\hspace{-0.2em}4}\xspace}
 \def\er #1 #2 { $#1 \pm #2$ }
 \def\bra #1 #2 #3 #4 { $#1 ^{+#2} _{-#3} \pm #4 $ }

\long\def\inst#1{\par\nobreak\kern 4pt\nobreak
    {\it #1}\par\vskip 10pt plus 3pt minus 3pt}

\begin{document}
{\pagestyle{empty}

\begin{flushright}
SLAC-PUB-\SLACPubNumber \\
\babar-PROC-\BABARPubYear/\BABARProcNumber \\
\today
\end{flushright}

\par\vskip 4cm

\begin{center}
\Large \bf Electro--Weak Penguin and Leptonic Decays in \babar
\end{center}
\bigskip

\begin{center}
\large 
F. Di Lodovico\\
Queen Mary, University of London \\
Physics Department, Mile End Road, London E1 4NS, UK \\
(for the \lbabar\ Collaboration)
\end{center}
\bigskip \bigskip

\begin{center}
\large \bf Abstract
\end{center}
Electro--weak penguin and leptonic decays provide an indirect probe
for physics beyond the Standard Model and contribute to the
determination of Standard Model parameters.
Copious quantities of \B\ mesons produced at the \B--Factories permit 
precision measurements of the electro--weak
penguin decays and searches for leptonic decays. 
We review the current experimental status of
$b \to s(d) \gamma$, $B^0\to D^{*0}\gamma$, $b \to s \ell^+\ell^- $ and finally $\btaunu$ decays
at \babar.

\vfill
\begin{center}
Contributed to the Proceedings of the \\
10th International Conference on \B--Physics at Hadron Machines, BEAUTY 2005,\\
6/20/2005---6/24/2005, Assisi (Perugia), Italy
\end{center}

\vspace{1.0cm}
\begin{center}
{\em Stanford Linear Accelerator Center, Stanford University, 
Stanford, CA 94309} \\ \vspace{0.1cm}\hrule\vspace{0.1cm}
Work supported in part by Department of Energy contract DE-AC02-76SF00515.
\end{center}

\section{Introduction}
Electro--weak penguin decays are flavor-changing neutral current 
transitions that are forbidden in the Standard Model
(SM) at tree level, but occur at loop level. 
Additional contributions to the electro--weak
penguin loops could arise from New Physics effects such as new gauge bosons,
charged Higgs bosons or supersymmetric particles. 
Study of electro--weak penguin decays constitutes an indirect probe for New Physics.
In addition, these decays are relevant to the determination of 
the CKM matrix elements $\mid V_{ub}\mid$ and $\mid V_{td}/V_{ts}\mid$ and
to the measurement of the photon polarization.

In the SM, leptonic decays proceed via quark annihilation
into a \Wp\ boson: $\bbar \u \to \Wp \to\ell^+ \nu_\ell $.
The study of the purely leptonic $\Bp \to \ell^+ \nu_\ell$ can act as an indirect 
probe for New Physics and can provide
sensitivity to poorly constrained SM parameters: 
$f_\B$, which is the \B\ decay constant, and $\mid V_{ub} \mid$, which is the
relevant CKM matrix element.

In this paper, recent results from $b \to s(d) \gamma$, $\Bz\to D^{*0}\gamma$, 
$b \to s \ell^+\ell^- $ and $\btaunu$ decays at \babar~\cite{Detector} 
are presented.
The data sample used in the analyses ranges between 88 and 232 millions of \BB\ events
collected at the \FourS\ peak.

%
%
\section{$b \to s \gamma$ final states}
Radiative decays involving the flavor--changing neutral current transition 
$b\to s$ are described in the SM primarily by 
a one--loop radiative penguin diagram containing a top quark and a $W$ boson. 
Additional contributions to the loop from New Physics, 
e.g. a charged Higgs boson or supersymmetric particles, could change the $b\to s\gamma$ 
rate~\cite{NewPhysics1,NewPhysics2,NewPhysics3,KaganNeubert,NewPhysics4},
the direct \CP\ asymmetry between $\b\to s\gamma$ and $\bbar\to \sbar\gamma$ decays~\cite{KNACP},
the direct asymmetry of the sum of $\b\to s\gamma$ and $\b\to \d\gamma$ decays~\cite{Hurth}
and the isospin asymmetry between charged and neutral mesons~\cite{IsospinKN}.
The photon energy spectrum provides access to the distribution 
function of the $b$ quark inside the \B\ meson, which is 
a crucial input in the extraction of \Vub\ from inclusive 
semileptonic $\B\to X_u\ell\nu$ 
measurements~\cite{ShapeFunction1,ShapeFunction2,ShapeFunction3,ShapeFunction4,ShapeFunction5,ShapeFunction6}. 
Finally, the decays $B\to K\pip\piz\gamma$, permit the measurement, given sufficient statistics, of the
photon polarization~\cite{Gronau02}. 
The SM predicts nearly complete left-handed polarization.

%
\subsection{$\B\to K^*\gamma$}

The exclusive $B \rightarrow K^* \gamma$ modes have been studied by 
\babar\ using a total 88 million of \BB\ pairs~\cite{kstargamma}.

We extract the final results utilizing kinematic constraints, resulting from the candidate \B\ meson reconstruction,
a neural network built using event-shape variables to reduce the continuum
background and a multi-dimensional extended maximum likelihood.

We find $\BR(\bkog)= (3.92 \pm 0.20({ stat }) \pm 0.24({  syst })) \times 10^{-5}$ and
$\BR(\bkpg) = (3.87 \pm 0.28({stat }) \pm 0.26({syst })) \times 10^{-5}$, consistent with the SM predictions.  Our
measurements also constrain the direct $\CP$ asymmetry to be ${\cal A}_{\CP} = - 0.013 \pm 0.036 (stat) \pm 0.010 (syst)$
and the isospin asymmetry to be $\Delta_{0+} = + 0.050 \pm 0.045 (stat) \pm 0.028 (syst)  \pm 0.024 (R^+/R^0) $.
The asymmetries are consistent with zero and are statistically limited.

\subsection{$B\to K\pi\pi\gamma$}
Using a total of 232 million \BB\ pairs, \babar\ has measured the partial branching fractions for $B\to
K\pi\pi\gamma$ in four decay channels in the range $\mkpipi <
1.8\gev$~\cite{kpppgamma}, including the $K\pi^+\pi^0\gamma$ channels important for
measuring the photon polarization. Table~\ref{results} shows the signal 
yields and the computed branching fraction for all the final states.
The results in the $K\pi^+\pi^-$
channels are consistent with the previous measurement~\cite{BelleKppg}. We
present first observations of decays in the $K\pi^+\pi^0$ channels.  
With higher statistics it will be possible to perform a measurement of the photon polarization,
but untangling the resonant contributions presents a challenge.

\begin{table}[htbp]
\caption{Results for $B\to K\pi\pi\gamma$, $m_{K\pi\pi}<1.8\ \hbox{GeV}$.  
The first column indicates the final states, the second column shows the signal yield 
and the corresponding statistical error, the
second column shows the branching fraction and the statistical and systematic
error, respectively.}
\begin{center}
\begin{tabular}{lcc}
\hline\hline
Channel  & Yield  & ${\cal B} (10^{-5})$ \\
\hline
\ccc                  &  $899 \pm 38$  & 2.95 $\pm$ 0.13 $\pm$ 0.19 \\
\ccn                  &  $572 \pm 31$  & 4.07 $\pm$ 0.22 $\pm$ 0.31 \\
$K^0\pi^+\pi^-\gamma$ &  $176 \pm 20$  & 1.85 $\pm$ 0.21 $\pm$ 0.12 \\
$K^0\pi^+\pi^0\gamma$ &  $164 \pm 15$  & 4.56 $\pm$ 0.42 $\pm$ 0.30 \\
\hline\hline
\end{tabular}
\end{center}
\label{results}
\end{table}
%
\subsection{Inclusive $b\to s \gamma$}
\babar\ has recently published a paper using 
a sum of exclusive final states~\cite{semiexcl} to study the inclusive $b\to s \gamma$ decay. 
\B\ candidates are obtained combining a $s$--quark hadronic system, $X_s$,
with a high energy photon. Kinematic constraints exploiting
the fact that the \B\ is produced from the \FourS\ decay
are used to suppress backgrounds.
Candidates with correctly reconstructed $X_s$ systems are treated as signal, whereas 
incorrectly reconstructed systems (``cross--feed'') are treated as background.
This method permits a measurement of 
the branching fraction as a function of the hadronic mass, $M(X_s)$. 
The $M(X_s)$ spectrum is converted into a high resolution photon energy spectrum 
using the kinematic relationship for the decay of a \B\ meson of mass $M_\B$:
$E_\gamma = {{M_\B^2 - M(X_s)^2}\over{2M_\B}}$.
where $E_\gamma$ is the photon energy in the \B\ rest frame 
which has a resolution of 1--5\,\mev. 
The photon energy spectrum is shown in Figure~\ref{fig:spectrum}.

\begin{figure}[htbp]
\begin{center}
 \begin{tabular}{c}
   \mbox{\includegraphics[width=10cm,height=8cm]{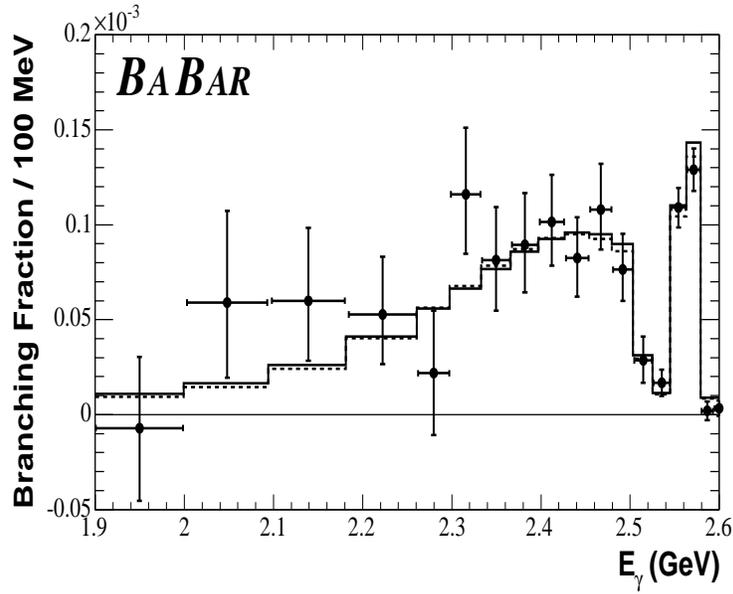}}
 \end{tabular}
\end{center}
\caption{The photon energy spectrum according to Ref.~\cite{semiexcl}. 
The data points are compared to theoretical predictions (histograms) obtained 
using the shape function (solid line) and
kinetic (dashed line) schemes.}
\label{fig:spectrum}
\end{figure}

We fit the photon energy spectrum to two recent theoretical predictions, 
one using a combination of the operator product expansion (OPE) coupled to soft 
collinear effective theory (SCET)~\cite{TheoryBF3,ShapeFunction4,ShapeFunction5,ShapeFunction6,N1,N2} 
in the shape function mass scheme,
and the other using a full OPE approach in the kinetic mass scheme~\cite{BBU}.
Figure~\ref{fig:spectrum} shown the best fit to the theoretical predictions, from which it can be seen 
that the spectrum is well described and the difference between the two schemes is small.
Results are $m_\b = 4.67\pm 0.07$~\gev, $\mu_\pi^2 = 0.16^{+0.10}_{-0.08}~\gev^2$ and 
$m_\b = 4.70^{\hspace{+0.4em}+\hspace{+0.4em}0.04}_{\hspace{+0.4em}-\hspace{+0.4em}0.08}$~\gev, 
$\mu_\pi^2 = 0.29^{+0.09}_{-0.04}~\gev^2$, where the errors are statistical and systematic combined,
for the two schemes, respectively.

Also \babar\ presented preliminary results from an analysis in which
the photon energy spectrum is measured without 
reconstructing the $X_s$ system, and backgrounds are suppressed using information from the rest of the event, and
tagging the other \B\ of the event. Details are given in Ref.~\cite{fullyincl}.
When measuring the $E_\gamma$ spectrum inclusively at the \FourS\ the
shape of the spectrum has a large contribution from 
the 50\,\mev\ calorimeter energy resolution, 
and from the motion of the \B\ meson in the \FourS\ rest frame.

We calculate the first and second moments of the photon spectrum 
for different minimum values of the photon energy.
The moments are in good agreement with predictions based 
on fits to the measured $\b\to c\ell\nu$ moments as it can be seen in Figure~\ref{fig:moments}.
\begin{figure}[htbp]
\begin{center}
\begin{tabular}{cc}
  \mbox{\includegraphics[width=8cm,height=6cm]{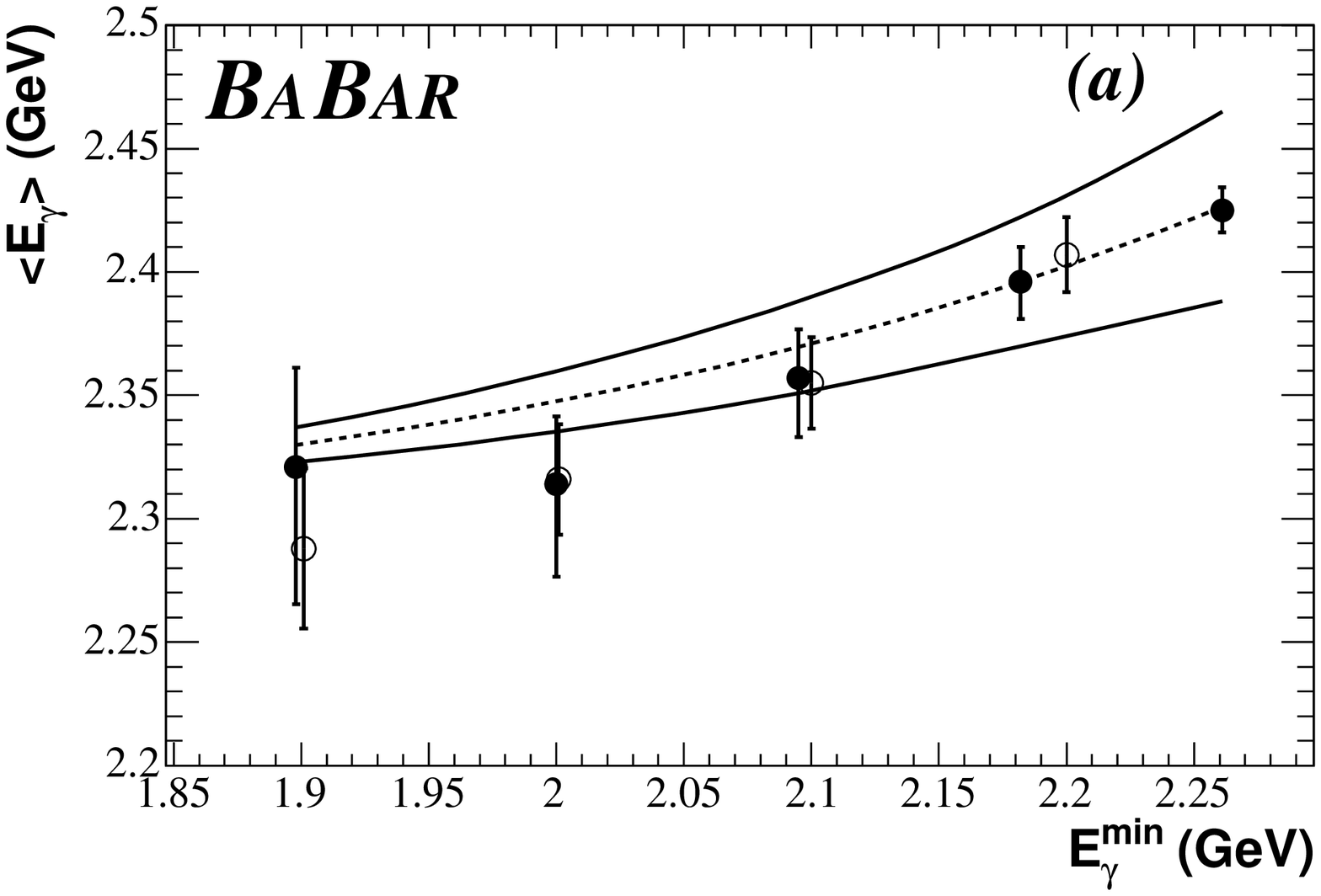}}&
  \mbox{\includegraphics[width=8cm,height=6cm]{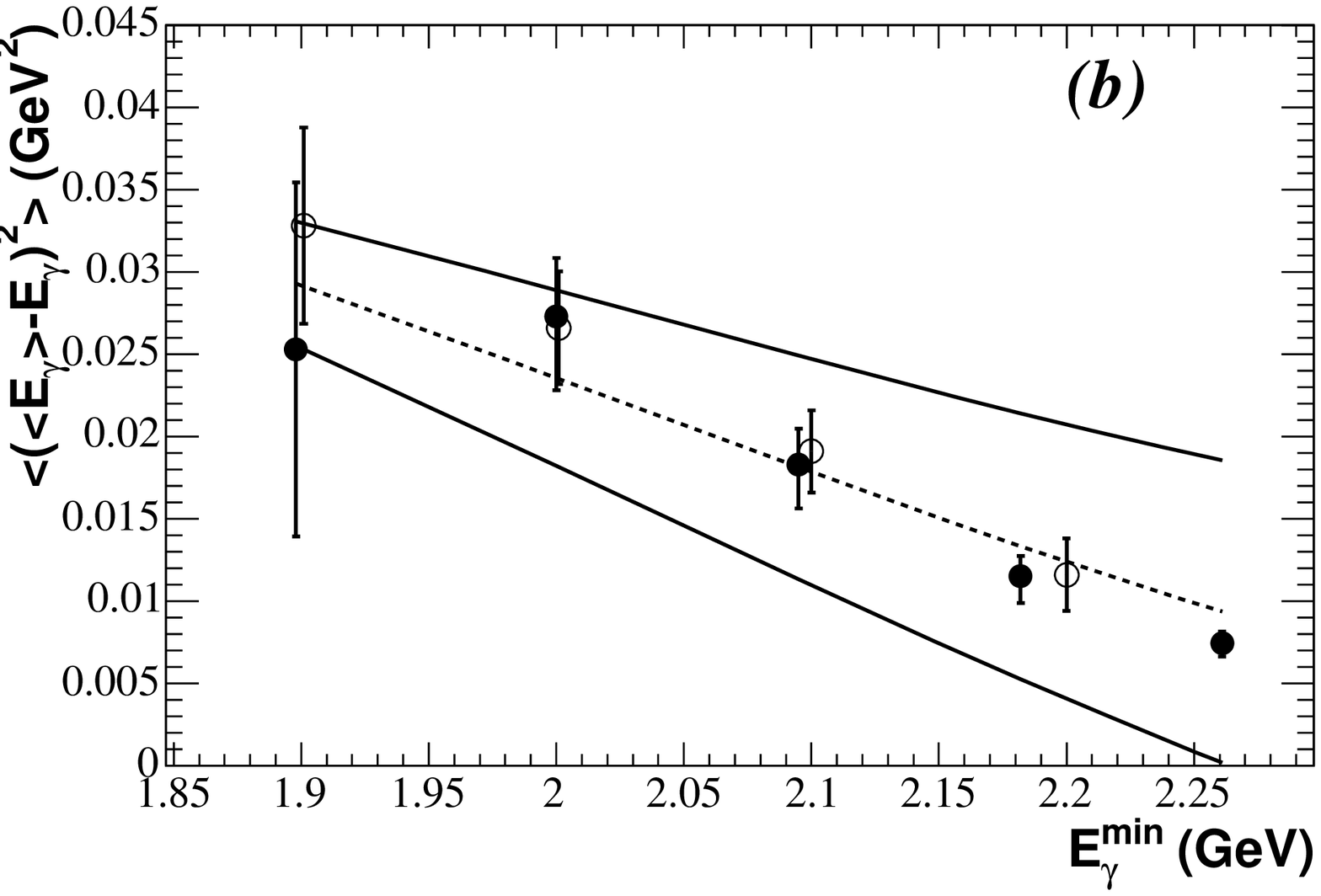}}\\
 \end{tabular}
\end{center}
\caption{First (a), and second (b) moments as a function of the minimum photon energy
according to Ref.~\cite{semiexcl} (solid dots) and Ref.~\cite{fullyincl} (empty dots).
The dotted lines show the predicted central values based on fits to
the $b\to c\ell\nu$ moments~\cite{Vcb}, and the solid lines the theoretical
uncertainties from the kinetic scheme~\cite{BBU,Uraltsev}.}
\label{fig:moments}
\end{figure}

The measured branching fraction for $E_{\gamma}>1.90\gev$ are
${\cal B}(\b\to s\gamma) = (3.27\pm 0.18 ^{+0.55 +0.04}_{-0.40 -0.09})\times 10^{-4}$ and
${\cal B}(\b\to s\gamma) = (3.67\pm 0.29 \pm 0.34 \pm 0.29)\times 10^{-4}$, where the errors are statistical,
systematic and theoretical, respectively,
for the analyses described in Refs.~\cite{semiexcl,fullyincl}. The measured branching fractions 
are consistent with the SM predictions. 

The direct \CP\ asymmetry between inclusive $\b\to s\gamma$
and $\bbar \to\sbar\gamma$ decays, expected to be less than 0.01 in the Standard Model, is measured to be
$A_{\CP}(\b\to s\gamma) = 0.025\pm 0.050(stat)\pm 0.015(syst)$~\cite{semiexclACP}, which is statistically 
limited and consistent with the SM predictions.
Looking at both $s$ and $d$ final states, the measured \CP\ asymmetry is 
$A_{\CP}(\b\to (s+d)\gamma) = -0.110\pm 0.115(stat)\pm 0.017(syst)$~\cite{fullyincl}, still statistically limited.

Finally, using the semi--inclusive method, we have made the first measurement of the isospin asymmetry 
between $\Bm\to X_{s\bar{u}}\gamma$ and $\Bzb\to X_{s\bar{d}}\gamma$. We obtain
$\Delta_{0-} = -0.006 \pm 0.058 \pm 0.009 \pm 0.024 $ which 
is consistent with zero within the experimental uncertainty, 
which is mainly statistical.

\section{$b \to \d \gamma$ final states}

Studies of the $b\to d \gamma $ decays for now focus primarily 
on searching for the exclusive process $\B\to \rho/\omega \gamma$.
Both inclusive and exclusive $b \rightarrow d \gamma$ decays are 
suppressed by $\mid V_{td} / V_{ts} \mid^2 \sim 0.04$ with respect to the corresponding
$b \rightarrow s \gamma$ modes.
The branching fraction is predicted to be in the range 
${\cal B} (B  \rightarrow \rho \gamma) = (0.9-2.7) \times 10^{-6}$
\cite{bosch}, while the \CP\ asymmetry is of the order of $10\%$ \cite{bosch}.

From the experimental point of view, $\B\to \rho(\omega) \gamma$ 
is more difficult than $\B\to K^*\gamma$ because the backgrounds are bigger since this mode is
CKM suppressed and $u \bar u, d \bar d$ continuum processes are enhanced
compared to $s \bar s$ continuum processes.

The upper limits at 90\%\ CL on the exclusive decays $B\to \rho(\omega)\gamma$ 
from \babar~\cite{BABARrhogamma}, which uses a neural network to 
suppress most of the continuum background and 
particle identification to veto kaons, are
0.4, 1.8 and 1.0$\times 10^{-6}$ on $\rho^0\gamma$, $\rho^+\gamma$ 
and $\omega\gamma$, respectively, using 211 million \BB\ events. Assuming isospin symmetry, this gives a 
combined limit ${\cal B}(B\to\rho\gamma)<1.2\times 10^{-6}$ (90~\% C.L.).
Using 386 million \BB\ decays, Belle observed a signal and measured a combined limit 
${\cal B}(B\to\rho\gamma) = (1.34^{+0.34}_{-0.31}(stat)^{+0.14}_{-0.10}(syst)) \times 10^{-6}$~\cite{Bellerhogamma},
in the region already excluded by \babar\ at 90\%\ C.L.
Updating the measurements using a larger statistical sample will help in understanding the difference among
the results obtained by the two Collaborations.

Of particular theoretical interested is the ratio ${\cal B}(\B\to\rho\gamma)$ 
to ${\cal}(\B\to K^*\gamma)$ as most of the theoretical uncertainty
cancels and can be used to determine the ratio $\mid V_{td} / V_{ts} \mid$.
Combining the results from \babar\ and Belle assuming isospin asymmetry, a branching fraction of
${\cal B} (B  \rightarrow \rho \gamma) =(0.94_{+0.25}^{-0.22})\times 10^{-6}$ is measured, from which
a constraint on the CKM elements $\mid V_{td} / V_{ts} \mid$ of
$0.18 \pm 0.03$ is obtained (see Ref.~\cite{UTfit} for details). 

%
%
\section{$\B^0\to D^{*0}\gamma$}
%
%
Within the Standard Model, the rare decay $\B^0\to D^{*0}\gamma$ is
dominated by the \W--boson exchange process.  
Similar \W--exchange transitions are present in other decays.  For
example, they contribute to the decay $\Bz\to\rho^0\g$ along with the
leading electromagnetic-penguin process.
The presence of an annihilation contribution in $\B\to\rho\gamma$ 
which is not present in $\B\to K^*\gamma$ decays and which is associated to a different CKM 
factor ($\sim \mid V^*_{ub}V_{ud}\mid$) represents a theoretical error in the extraction 
of $\mid V_{td} / V_{ts} \mid$.

Using 88 million \BB\ pairs, \babar\ sets an 
upper limit on the branching fraction of ${\cal B}(\B^0\to D^{*0}\gamma) < 2.5\times 10^{-5}$ at
the 90\%\ confidence level~\cite{Dstar0gamma}, twice as stringent as the previous limit.

%
%
\section{$b\to s \ell^+\ell^- $ final states}
%
%
The rare decay \BtoXsll, which proceeds through the
$b \to s \ell^{+} \ell^{-}$ transition, 
is forbidden at lowest order in the SM
but is allowed at higher order via electroweak penguin and $W$-box
diagrams.
New Physics can appear in the loops and change the rate~\cite{Ali02,Hurth03},
the asymmetry~\cite{bib:krugercp} and the kinematic distributions.

\babar\ has measured the branching fractions and direct $CP$ asymmetries
$A_{CP}$ of the rare decays \modekavgll and \modekstll, using a total of 229 million \BB\ pairs~\cite{kll}.

We select events that include two oppositely charged electrons or muons,
a kaon candidate and, for the $B\to K^{*}\ell^+\ell^-$ modes, a $\pi^{\pm}$
candidate that, when combined with a kaon candidate, forms a $K^*$ candidate.
Combinatorial background from continuum processes is suppressed using a
Fisher discriminant.
Background \B\ decays to charmonium are suppressed 
excluding dilepton pairs consistent with the
$J/\psi$ and $\psi(2S)$ masses.
Fits to the kinematic variables of the reconstructed \B\ mesons are performed to 
extract the signal events.

We find the (lepton-flavor--averaged, \B-charge--averaged) branching fractions
${\cal B}(\modekavgll)=(0.34\pm 0.07 \pm 0.03)\times 10^{-6}$ and
${\cal B}(\modekstll)=(0.78^{+0.19}_{-0.17}\pm 0.12)\times 10^{-6},$
consistent with the SM predictions for these modes.  
We measure the direct $CP$ asymmetries
$A_{CP}(\modekll)=0.08\pm0.22\pm 0.11$ and
$A_{CP}(\modekstll)=-0.03\pm0.23\pm 0.12$
where the systematic uncertainty is dominated by the unknown asymmetry in
the peaking backgrounds.

Studying separately electron and muon final states, we find the ratio of muon to electron branching
fractions over the full range of $q^2$ to be
$R_{K} = 1.06\pm0.48\pm 0.05$ and
$R_{K^*} = 0.93\pm0.46\pm 0.06$
where these are expected in the Standard Model to be 1.00 and 0.75,
respectively, with small theoretical uncertainties.

All measurements are consistent with the SM predictions and are statistically limited.

\babar\ also observed the inclusive decay $B\to X_s \ell \ell$. Details can be found in Ref.~\cite{Xsll}.

%
%
\section{$\btaunu$}
We search for the rare leptonic decay $\btaunu$ in a sample of
$232 \times 10^6$ \BB\ pairs~\cite{btaunu}. 
This decay is helicity favoured with respect to the corresponding decays with electrons and
muons.
Signal events are selected by 
examining the properties of the \B\ meson
recoiling against the semileptonic decay \btodszlnu.
We find no evidence for a signal and set an upper limit on the branching
fraction of $\BRbtaunu < 2.8 \times 10^{-4}$ at the 90\% confidence level.
We combine this result
with a previous, statistically
independent \babar\ search for \btaunu\ to give an upper limit of
$\mathcal{B}(\btaunu) < 2.6 \times 10^{-4}$ at the 90\% confidence level.
These result represents the most stringent
limit on $\btaunu$ reported to date.
Using limits on $\B\to\tau\nu$ from both \babar\ and Belle, a value
$ f_\B = 0.178 \pm 0.062 \gev $ is obtained from fits to the
unitarity triangle (see Ref.~\cite{UTfit} for details).
%
%
\section{Conclusions}
A review of recent experimental results on electro--weak and leptonic decays 
is presented. In particular, we have focussed on the new results on the 
$b\to s (d) \gamma$, $\B\to D^{*0}\gamma$, $b\to s \ell \ell $ and $\B \to \tau \nu$
decays.

%
%
\section{Acknowledgements}
We are grateful to Steve Playfer for several useful discussions.
We would like to thank the organizers of BEAUTY 2005 for the invitation and for having 
organized a very interesting conference.


\end{document}